\documentclass[prl,twocolumn]{revtex4}
\usepackage{amsmath}
\usepackage{amsfonts}
\usepackage{mathrsfs}
\usepackage{amssymb}
\usepackage{bm}
\usepackage{bbm}
\usepackage{slashed}
\usepackage{rotating}
\usepackage{mciteplus}

\begin{document}

\newlength{\wdo}
\newcommand{\stroke}[1]{{$#1$}%
\settowidth{\wdo}{${#1}$} {\kern-\wdo}%
\partialvartstrokedint}

\newenvironment{ibox}[1]%
{\vskip 1.0em
\framebox[\columnwidth][r]{%
\begin{minipage}[c]{\columnwidth}%
\vspace{-1.0em}%
#1%
\end{minipage}}}
{\vskip 1.0em}

\newcommand{\iboxed}[1]{%
\vskip 1.0em
\framebox[\columnwidth][r]{%
\begin{minipage}[c]{\columnwidth}%
\vspace{-1.0em}
#1%
\end{minipage}}
\vskip 1.0em}

\newcommand{\fitbox}[2]{%
\vskip 1.0em
\begin{flushright}
\framebox[{#1}][r]{%
\begin{minipage}[c]{\columnwidth}%
\vspace{-1.0em}
#2%
\end{minipage}}
\end{flushright}
\vskip 1.0em}

\newcommand{\iboxeds}[1]{%
\vskip 1.0em
\begin{equation}
\fbox{%
\begin{minipage}[c]{1mm}%
\vspace{-1.0em}
#1%
\end{minipage}}
\end{equation}
\vskip 1.0em}

\def\Xint#1{\mathchoice
   {\XXint\displaystyle\textstyle{#1}}%
   {\XXint\textstyle\scriptstyle{#1}}%
   {\XXint\scriptstyle\scriptscriptstyle{#1}}%
   {\XXint\scriptscriptstyle\scriptscriptstyle{#1}}%
   \!\int}
\def\XXint#1#2#3{{\setbox0=\hbox{$#1{#2#3}{\int}$}
     \vcenter{\hbox{$#2#3$}}\kern-.5\wd0}}
\def\ddashint{\Xint=}
\def\dashint{\Xint-}

\def\mathbi#1{\textbf{\em #1}}

\newcommand{\alps}{\ensuremath{\alpha_s}}
\newcommand{\qbar}{\bar{q}}
\newcommand{\ubar}{\bar{u}}
\newcommand{\dbar}{\bar{d}}
\newcommand{\sbar}{\bar{s}}
\newcommand{\beq}{\begin{equation}}
\newcommand{\eeq}{\end{equation}}
\newcommand{\beqa}{\begin{eqnarray}}
\newcommand{\eeqa}{\end{eqnarray}}
\newcommand{\gs}{g_{\pi NN}}
\newcommand{\gw}{f_\pi}
\newcommand{\mq}{m_Q}
\newcommand{\mn}{m_N}
\newcommand{\mpi}{m_\pi}
\newcommand{\mrho}{m_\rho}
\newcommand{\momg}{m_\omega}
\newcommand{\bb}{\langle}
\newcommand{\kb}{\rangle}
\newcommand{\xvec}{\mathbf{x}}
\newcommand{\st}{\ensuremath{\sqrt{\sigma}}}
\newcommand{\Bvec}{\mathbf{B}}
\newcommand{\rvec}{\mathbf{r}}
\newcommand{\kvec}{\mathbf{k}}
\newcommand{\pvec}{\mathbf{p}}
\newcommand{\Pvec}{\mathbf{P}}
\newcommand{\bvec}[1]{\ensuremath{\mathbf{#1}}}
\newcommand{\bra}[1]{\ensuremath{\bb#1|}}
\newcommand{\ket}[1]{\ensuremath{|#1\kb}}
\newcommand{\gft}{\ensuremath{\gamma_{FT}}}
\newcommand{\gfv}{\ensuremath{\gamma_5}}
\newcommand{\bfalp}{\ensuremath{\bm{\alpha}}}
\newcommand{\bfbeta}{\ensuremath{\bm{\beta}}}
\newcommand{\bfeps}{\ensuremath{\bm{\epsilon}}}
\newcommand{\lag}{{\lambda_\gamma}}
\newcommand{\lao}{{\lambda_\omega}}
\newcommand{\lN}{\lambda_N}
\newcommand{\lM}{\lambda_M}
\newcommand{\lB}{\lambda_B}
\newcommand{\epslag}{\ensuremath{\bm{\epsilon}_{\lag}}}
\newcommand{\bfept}{\ensuremath{\tilde{\bm{\epsilon}}}}
\newcommand{\bfgam}{\ensuremath{\bm{\gamma}}}
\newcommand{\bfnab}{\ensuremath{\bm{\nabla}}}
\newcommand{\bflambda}{\ensuremath{\bm{\lambda}}}
\newcommand{\bfmu}{\ensuremath{\bm{\mu}}}
\newcommand{\bfphi}{\ensuremath{\bm{\phi}}}
\newcommand{\bfvphi}{\ensuremath{\bm{\varphi}}}
\newcommand{\bfpi}{\ensuremath{\bm{\pi}}}
\newcommand{\bfsig}{\ensuremath{\bm{\sigma}}}
\newcommand{\bftau}{\ensuremath{\bm{\tau}}}
\newcommand{\bfpsi}{\ensuremath{\bm{\psi}}}
\newcommand{\bfdelta}{\ensuremath{\bm{\delta}}}
\newcommand{\bfrho}{\ensuremath{\bm{\rho}}}
\newcommand{\bfth}{\ensuremath{\bm{\theta}}}
\newcommand{\bfchi}{\ensuremath{\bm{\chi}}}
\newcommand{\bfxi}{\ensuremath{\bm{\xi}}}
\newcommand{\bfR}{\ensuremath{\bvec{R}}}
\newcommand{\bfP}{\ensuremath{\bvec{P}}}
\newcommand{\bfJ}{{\mathbi{J}}}
\newcommand{\bfj}{{\mathbi{j}}}
\newcommand{\Rcm}{\ensuremath{\bvec{R}_{CM}}}
\newcommand{\spup}{\uparrow}
\newcommand{\spd}{\downarrow}
\newcommand{\up}{\uparrow}
\newcommand{\dn}{\downarrow}
\newcommand{\hbarom}{\frac{\hbar^2}{m_Q}}
\newcommand{\half}{\ensuremath{\frac{1}{2}}}
\newcommand{\thalf}{\ensuremath{\frac{3}{2}}}
\newcommand{\fhalf}{\ensuremath{\frac{5}{2}}}
\newcommand{\shalf}{\ensuremath{{\tfrac{1}{2}}}}
\newcommand{\sqtr}{\ensuremath{{\tfrac{1}{4}}}}
\newcommand{\sphalf}{\ensuremath{\genfrac{}{}{0pt}{1}{+}{}\!\tfrac{1}{2}}}
\newcommand{\smhalf}{\ensuremath{\genfrac{}{}{0pt}{1}{-}{}\!\tfrac{1}{2}}}
\newcommand{\sthalf}{\ensuremath{{\tfrac{3}{2}}}}
\newcommand{\spthalf}{\ensuremath{{\tfrac{+3}{2}}}}
\newcommand{\smthalf}{\ensuremath{{\tfrac{-3}{2}}}}
\newcommand{\sfhalf}{{\tfrac{5}{2}}}
\newcommand{\third}{{\frac{1}{3}}}
\newcommand{\tthird}{{\frac{2}{3}}}
\newcommand{\sthird}{{\tfrac{1}{3}}}
\newcommand{\stthird}{{\tfrac{2}{3}}}
\newcommand{\vnn}{\ensuremath{\hat{v}_{NN}}}
\newcommand{\vij}{\ensuremath{\hat{v}_{ij}}}
\newcommand{\vik}{\ensuremath{\hat{v}_{ik}}}
\newcommand{\argonne}{\ensuremath{v_{18}}}
\newcommand{\lqcd}{\ensuremath{\mathcal{L}_{QCD}}}
\newcommand{\lqed}{\ensuremath{\mathscr{L}_{QED}}}
\newcommand{\lgf}{\ensuremath{\mathcal{L}_g}}
\newcommand{\lqm}{\ensuremath{\mathcal{L}_q}}
\newcommand{\lqg}{\ensuremath{\mathcal{L}_{qg}}}
\newcommand{\nn}{\ensuremath{N\!N}}
\newcommand{\nnn}{\ensuremath{N\!N\!N}}
\newcommand{\qq}{\ensuremath{qq}}
\newcommand{\qqq}{\ensuremath{qqq}}
\newcommand{\qqb}{\ensuremath{q\bar{q}}}
\newcommand{\hpnd}{\ensuremath{H_{\pi N\Delta}}}
\newcommand{\hpqq}{\ensuremath{H_{\pi qq}}}
\newcommand{\hpqqa}{\ensuremath{H^{(a)}_{\pi qq}}}
\newcommand{\hpqqe}{\ensuremath{H^{(e)}_{\pi qq}}}
\newcommand{\hint}{\ensuremath{H_{\rm int}}}
\newcommand{\fpnn}{\ensuremath{f_{\pi\! N\!N}}}
\newcommand{\fenn}{\ensuremath{f_{\eta\! N\!N}}}
\newcommand{\gsnn}{\ensuremath{g_{\sigma\! N\!N}}}
\newcommand{\gpnn}{\ensuremath{g_{\pi\! N\!N}}}
\newcommand{\fpnd}{\ensuremath{f_{\pi\! N\!\Delta}}}
\newcommand{\grpg}{\ensuremath{g_{\rho\pi\gamma}}}
\newcommand{\gopg}{\ensuremath{g_{\omega\pi\gamma}}}
\newcommand{\fmqq}{\ensuremath{f_{M\! qq}}}
\newcommand{\gmqq}{\ensuremath{g_{M\! qq}}}
\newcommand{\fpqq}{\ensuremath{f_{\pi qq}}}
\newcommand{\gpqq}{\ensuremath{g_{\pi qq}}}
\newcommand{\feqq}{\ensuremath{f_{\eta qq}}}
\newcommand{\gonn}{\ensuremath{g_{\omega N\!N}}}
\newcommand{\gonna}{\ensuremath{g^t_{\omega N\!N}}}
\newcommand{\grnn}{\ensuremath{g_{\rho N\!N}}}
\newcommand{\gopr}{\ensuremath{g_{\omega\pi\rho}}}
\newcommand{\grnp}{\ensuremath{g_{\rho N\!\pi}}}
\newcommand{\grpp}{\ensuremath{g_{\rho\pi\pi}}}
\newcommand{\Lpnn}{\ensuremath{\Lambda_{\pi\! N\! N}}}
\newcommand{\Lonn}{\ensuremath{\Lambda_{\omega N\! N}}}
\newcommand{\Lonna}{\ensuremath{\Lambda^t_{\omega N\! N}}}
\newcommand{\Lrnn}{\ensuremath{\Lambda_{\rho N\! N}}}
\newcommand{\Lopr}{\ensuremath{\Lambda_{\omega\pi\rho}}}
\newcommand{\Lrpp}{\ensuremath{\Lambda_{\rho\pi\pi}}}
\newcommand{\getaqq}{\ensuremath{g_{\eta qq}}}
\newcommand{\fsqq}{\ensuremath{f_{\sigma qq}}}
\newcommand{\gsqq}{\ensuremath{g_{\sigma qq}}}
\newcommand{\piqq}{\ensuremath{{\pi\! qq}}}
\newcommand{\ylm}{\ensuremath{Y_\ell^m}}
\newcommand{\ylmc}{\ensuremath{Y_\ell^{m*}}}
\newcommand{\ebh}[1]{\hat{\bvec{e}}_{#1}}
\newcommand{\kbh}{\hat{\bvec{k}}}
\newcommand{\nbh}{\hat{\bvec{n}}}
\newcommand{\pvbh}{\hat{\bvec{p}}}
\newcommand{\qbh}{\hat{\bvec{q}}}
\newcommand{\Xbh}{\hat{\bvec{X}}}
\newcommand{\rbh}{\hat{\bvec{r}}}
\newcommand{\xbh}{\hat{\bvec{x}}}
\newcommand{\ybh}{\hat{\bvec{y}}}
\newcommand{\zbh}{\hat{\bvec{z}}}
\newcommand{\betabh}{\hat{\bfbeta}}
\newcommand{\tbh}{\hat{\bfth}}
\newcommand{\pbh}{\hat{\bfvphi}}
\newcommand{\dt}{\Delta\tau}
\newcommand{\kmag}{|\bvec{k}|}
\newcommand{\pmag}{|\bvec{p}|}
\newcommand{\qmag}{|\bvec{q}|}
\newcommand{\oas}{\ensuremath{\mathcal{O}(\alpha_s)}}
\newcommand{\vtxb}{\ensuremath{\Lambda_\mu(p',p)}}
\newcommand{\vtxp}{\ensuremath{\Lambda^\mu(p',p)}}
\newcommand{\pwqp}{e^{i\bvec{q}\cdot\bvec{r}}}
\newcommand{\pwqm}{e^{-i\bvec{q}\cdot\bvec{r}}}
\newcommand{\gsa}[1]{\ensuremath{\bb#1\kb_0}}
\newcommand{\oer}[1]{\mathcal{O}\left(\frac{1}{\qmag^{#1}}\right)}
\newcommand{\nub}[1]{\overline{\nu^{#1}}}
\newcommand{\epf}{E_\bvec{p}}
\newcommand{\epfp}{E_{\bvec{p}'}}
\newcommand{\eka}{E_{\alpha\kappa}}
\newcommand{\ekaq}{(E_{\alpha\kappa})^2}
\newcommand{\ekap}{E_{\alpha'\kappa}}
\newcommand{\ekpa}{E+{\alpha\kappa_+}}
\newcommand{\ekma}{E_{\alpha\kappa_-}}
\newcommand{\ekp}{E_{\kappa_+}}
\newcommand{\ekm}{E_{\kappa_-}}
\newcommand{\ekpap}{E_{\alpha'\kappa_+}}
\newcommand{\ekmap}{E_{\alpha'\kappa_-}}
\newcommand{\yjm}[1]{\mathcal{Y}_{jm}^{#1}}
\newcommand{\ysa}[3]{\mathcal{Y}_{#1,#2}^{#3}}
\newcommand{\yss}[2]{\mathcal{Y}_{#1}^{#2}}
\newcommand{\Dj}{\ensuremath{\mathscr{D}}}
\newcommand{\ysc}{\tilde{y}}
\newcommand{\enm}{\varepsilon_{NM}}
\newcommand{\Scg}[6]
	{\ensuremath{S^{#1}_{#4}\:\vphantom{S}^{#2}_{#5}
 	 \:\vphantom{S}^{#3}_{#6}\,}}
\newcommand{\Kmat}[6]
	{\ensuremath{K\left[\begin{array}{ccc} 
	#1 & #2 & #3 \\ #4 & #5 & #6\end{array}\right]}}
\newcommand{\irt}{\ensuremath{\frac{1}{\sqrt{2}}}}
\newcommand{\sirt}{\ensuremath{\tfrac{1}{\sqrt{2}}}}
\newcommand{\irth}{\ensuremath{\frac{1}{\sqrt{3}}}}
\newcommand{\sirth}{\ensuremath{\tfrac{1}{\sqrt{3}}}}
\newcommand{\irs}{\ensuremath{\frac{1}{\sqrt{6}}}}
\newcommand{\sirs}{\ensuremath{\tfrac{1}{\sqrt{6}}}}
\newcommand{\tors}{\ensuremath{\frac{2}{\sqrt{6}}}}
\newcommand{\stors}{\ensuremath{\tfrac{2}{\sqrt{6}}}}
\newcommand{\rtoth}{\ensuremath{\sqrt{\frac{2}{3}}}}
\newcommand{\rthot}{\ensuremath{\frac{\sqrt{3}}{2}}}
\newcommand{\ithrt}{\ensuremath{\frac{1}{3\sqrt{2}}}}
\newcommand{\Tg}{\ensuremath{\mathsf{T}}}
\newcommand{\irrep}[1]{\ensuremath{\mathbf{#1}}}
\newcommand{\cirrep}[1]{\ensuremath{\overline{\mathbf{#1}}}}
\newcommand{\Fij}{\ensuremath{\hat{F}_{ij}}}
\newcommand{\Fqij}{\ensuremath{\hat{F}^{(qq)}_{ij}}}
\newcommand{\Fsij}{\ensuremath{\hat{F}^{(qs)}_{ij}}}
\newcommand{\Opij}{\mathcal{O}^p_{ij}}
\newcommand{\fpij}{f_p(r_{ij})}
\newcommand{\titj}{\bftau_i\cdot\bftau_j}
\newcommand{\sisj}{\bfsig_i\cdot\bfsig_j}
\newcommand{\Sij}{S_{ij}}
\newcommand{\LS}{\bvec{L}_{ij}\cdot\bvec{S}_{ij}}
\newcommand{\TT}{\Tg_i\cdot\Tg_j}
\newcommand{\chet}{\ensuremath{\chi ET}}
\newcommand{\chpt}{\ensuremath{\chi PT}}
\newcommand{\chsy}{\ensuremath{\chi\mbox{symm}}}
\newcommand{\lchi}{\ensuremath{\Lambda_\chi}}
\newcommand{\lcon}{\ensuremath{\Lambda_{QCD}}}
\newcommand{\dcpsi}{\ensuremath{\bar{\psi}}}
\newcommand{\dcbfpsi}{\ensuremath{\bar{\bfpsi}}}
\newcommand{\dc}[1]{\ensuremath{\overline{#1}}}
\newcommand{\dcpsip}{\ensuremath{\bar{\psi}^{(+)}}}
\newcommand{\psip}{\ensuremath{{\psi}^{(+)}}}
\newcommand{\dcpsim}{\ensuremath{\bar{\psi}^{(-)}}}
\newcommand{\psim}{\ensuremath{{\psi}^{(-)}}}
\newcommand{\llo}{\ensuremath{\mathcal{L}^{(0)}_{\chet}}}
\newcommand{\lchet}{\ensuremath{\mathcal{L}_{\chi}}}
\newcommand{\hchet}{\ensuremath{\mathcal{H}_{\chi}}}
\newcommand{\Hd}{\ensuremath{\mathcal{H}}}
\newcommand{\Dmu}{\ensuremath{\mathcal{D}_\mu}}
\newcommand{\Dsl}{\ensuremath{\slashed{\mathcal{D}}}}
\newcommand{\comm}[2]{\ensuremath{\left[#1,#2\right]}}
\newcommand{\acomm}[2]{\ensuremath{\left\{#1,#2\right\}}}
\newcommand{\ev}[1]{\ensuremath{\bb\hat{#1}\kb}}
\newcommand{\exv}[1]{\ensuremath{\bb{#1}\kb}}
\newcommand{\evt}[1]{\ensuremath{\bb{#1}(\tau)\kb}}
\newcommand{\evm}[1]{\ensuremath{\bb{#1}\kb_M}}
\newcommand{\evv}[1]{\ensuremath{\bb{#1}\kb_V}}
\newcommand{\ovl}[2]{\ensuremath{\bb{#1}|{#2}\kb}}
\newcommand{\pd}{\partial}
\newcommand{\pnpd}[2]{\frac{\partial{#1}}{\partial{#2}}}
\newcommand{\pppd}[1]{\frac{\partial{\hphantom{#1}}}{\partial{#1}}}
\newcommand{\fnfd}[2]{\frac{\delta{#1}}{\delta{#2}}}
\newcommand{\rfnfd}[2]{\frac{\bfdelta{#1}}{\bfdelta{#2}}}
\newcommand{\fdfd}[1]{\frac{\delta}{\delta{#1}}}
\newcommand{\rfdfd}[1]{\frac{\overleftarrow{\delta}}{\delta{#1}}}
\newcommand{\plmu}{\partial_\mu}
\newcommand{\plnu}{\partial_\nu}
\newcommand{\pumu}{\partial^\mu}
\newcommand{\punu}{\partial^\nu}
\newcommand{\mcdf}{\delta^{(4)}(p_f-p_i-q)}
\newcommand{\ecdf}{\delta(E_f-E_i-\nu)}
\newcommand{\tr}{\mbox{Tr }}
\newcommand{\lxr}{\ensuremath{SU(2)_L\times SU(2)_R}}
\newcommand{\gV}[2]{\ensuremath{(\gamma^{-1})^{#1}_{\hphantom{#1}{#2}}}}
\newcommand{\gVd}[2]{\ensuremath{\gamma^{#1}_{\hphantom{#1}{#2}}}}
\newcommand{\LpV}[1]{\ensuremath{\Lambda^{#1}V}}
\newcommand{\hatH}{\ensuremath{\hat{H}}}
\newcommand{\hath}{\ensuremath{\hat{h}}}
\newcommand{\eht}{\ensuremath{e^{-\tau\hat{H}}}}
\newcommand{\ehdt}{\ensuremath{e^{-\Delta\tau\hat{H}}}}
\newcommand{\ehtm}{\ensuremath{e^{-\tau(\hat{H}-E_V)}}}
\newcommand{\ehdtm}{\ensuremath{e^{-\Delta\tau(\hat{H}-E_V)}}}
\newcommand{\Oop}{\ensuremath{\mathcal{O}}}
\newcommand{\Jop}{\ensuremath{\mathcal{J}}}
\newcommand{\Gop}{\ensuremath{\hat{\mathcal{G}}}}
\newcommand{\SU}[1]{\ensuremath{SU({#1})}}
\newcommand{\U}[1]{\ensuremath{U({#1})}}
\newcommand{\proj}[1]{\ensuremath{\ket{#1}\bra{#1}}}
\newcommand{\su}[1]{\ensuremath{\mathfrak{su}({#1})}}
\newcommand{\ip}[2]{\ensuremath{\bvec{#1}\cdot\bvec{#2}}}
\newcommand{\norm}[1]{\ensuremath{\left| #1\right|^2}}
\newcommand{\rnorm}[1]{\ensuremath{\lvert #1\rvert}}
\newcommand{\pid}{\left(\begin{array}{cc} 1 & 0 \\ 0 & 1\end{array}\right)}
\newcommand{\psx}{\left(\begin{array}{cc} 0 & 1 \\ 1 & 0\end{array}\right)}
\newcommand{\psy}{\left(\begin{array}{cc} 0 & -i \\ i & 0\end{array}\right)}
\newcommand{\psz}{\left(\begin{array}{cc} 1 & 0 \\ 0 & -1\end{array}\right)}
\newcommand{\ua}{\uparrow}
\newcommand{\da}{\downarrow}
\newcommand{\deln}{\delta_{i_1 i_2\ldots i_n}}
\newcommand{\GabRR}{G_{\alpha\beta}(\bfR,\bfR')}
\newcommand{\GRR}{G(\bfR,\bfR')}
\newcommand{\GfRR}{G_0(\bfR,\bfR')}
\newcommand{\GRiR}{G(\bfR_i,\bfR_{i-1})}
\newcommand{\GRRs}[2]{G(\bfR_{#1},\bfR_{#2})}
\newcommand{\Gdgn}{\Gamma_{\Delta,\gamma N}}
\newcommand{\Gdgnb}{\overline\Gamma_{\Delta,\gamma N}}
\newcommand{\GJT}{\Gamma_{LS}^{JT}(k)}
\newcommand{\GJTa}[2]{\Gamma^{#1}_{#2}}
\newcommand{\GtwJTa}[2]{\tilde{\Gamma}_{#1}^{#2}}
\newcommand{\Gtw}{\tilde{\Gamma}}
\newcommand{\Gbar}{\overline{\Gamma}}
\newcommand{\Gtil}{\tilde{\Gamma}}
\newcommand{\Gpndb}{\overline{\Gamma}_{\pi N,\Delta}}
\newcommand{\GbNgn}{{\overline{\Gamma}}_{N^*,\gamma N}}
\newcommand{\GNgn}{\Gamma_{N^*,\gamma N}}
\newcommand{\GbNmb}{{\overline{\Gamma}}_{N^*,MB}}
\newcommand{\Lg}[2]{\ensuremath{L^{#1}_{\hphantom{#1}{#2}}}}
\newcommand{\psik}{\ensuremath{\left(\begin{matrix}\psi_1 \\ \psi_2\end{matrix}\right)}}
\newcommand{\psib}{\ensuremath{\left(\begin{matrix}\psi^*_1&\psi^*_2\end{matrix}\right)}}
\newcommand{\Gf}{\ensuremath{\frac{1}{E-H_0}}}
\newcommand{\Gv}{\ensuremath{\frac{1}{E-H_0-\vnres}}}
\newcommand{\Gx}{\ensuremath{\frac{1}{E-H_0-V}}}
\newcommand{\Gex}{\ensuremath{\mathcal{G}}}
\newcommand{\Gfpm}{\ensuremath{\frac{1}{E-H_0\pm i\epsilon}}}
\newcommand{\vres}{v_R}
\newcommand{\vnres}{v}
\newcommand{\tpz}{\ensuremath{^3P_0}}
\newcommand{\tres}{t_R}
\newcommand{\tsr}{t^R}
\newcommand{\tsnr}{t^{NR}}
\newcommand{\trest}{\tilde{t}_R}
\newcommand{\tnres}{t}
\newcommand{\Pt}{P_{12}}
\newcommand{\Sz}{\ket{S_0}}
\newcommand{\Sa}{\ket{S^{(-1)}_1}}
\newcommand{\Sb}{\ket{S^{(0)}_1}}
\newcommand{\Sc}{\ket{S^{(+1)}_1}}
\newcommand{\sbasis}{\ket{s_1 s_2; m_1 m_2}}
\newcommand{\Sbasis}{\ket{s_1 s_2; S M}}
\newcommand{\sket}[2]{\ket{{#1}\,{#2}}}
\newcommand{\sbra}[2]{\bra{{#1}\,{#2}}}
\newcommand{\psmket}{\ket{\bvec{p};s\,m}}
\newcommand{\cket}{\ket{\bvec{p};s_1 s_2\,m_1 m_2}}
\newcommand{\hket}{\ket{\bvec{p};s_1 s_2\,\lambda_1\lambda_2}}
\newcommand{\hkets}{\ket{s\,\lambda}}
\newcommand{\phkets}{\ket{\bvec{p};s\,\lambda}}
\newcommand{\klsjm}{\ket{p;\ell s; j m}}
\newcommand{\pq}{\bvec{p}_q}
\newcommand{\pqb}{\bvec{p}_{\qbar}}
\newcommand{\mps}[1]{\frac{d^3{#1}}{(2\pi)^{3/2}}}
\newcommand{\mpsf}[1]{\frac{d^3{#1}}{(2\pi)^{3}}}
\newcommand{\du}[1]{u_{\bvec{#1},s}}
\newcommand{\dv}[1]{v_{\bvec{#1},s}}
\newcommand{\cdu}[1]{\overline{u}_{\bvec{#1},s}}
\newcommand{\cdv}[1]{\overline{v}_{\bvec{#1},s}}
\newcommand{\dus}[2]{u_{\bvec{#1},{#2}}}
\newcommand{\dvs}[2]{v_{\bvec{#1},{#2}}}
\newcommand{\cdus}[2]{\overline{u}_{\bvec{#1},{#2}}}
\newcommand{\cdvs}[2]{\overline{v}_{\bvec{#1},{#2}}}
\newcommand{\bop}[1]{b_{\bvec{#1},s}}
\newcommand{\dop}[1]{d_{\bvec{#1},s}}
\newcommand{\bops}[2]{b_{\bvec{#1},{#2}}}
\newcommand{\dops}[2]{d_{\bvec{#1},{#2}}}
\newcommand{\mev}{\mbox{ MeV}}
\newcommand{\gev}{\mbox{ GeV}}
\newcommand{\fmi}{\mbox{ fm}}
\newcommand{\M}{\mathcal{M}}
\newcommand{\Smat}{\mathcal{S}}
\newcommand{\JLSTh}{JLST\lambda}
\newcommand{\Tpg}{T_{\pi N,\gamma N}}
\newcommand{\tpg}{t_{\pi N,\gamma N}}
\newcommand{\vmbmb}{\ensuremath{v_{M'B',MB}}}
\newcommand{\tmbgn}{\ensuremath{t_{MB,\gamma N}}}
\newcommand{\Tonon}{\ensuremath{T_{\omega N,\omega N}}}
\newcommand{\tonon}{\ensuremath{t_{\omega N,\omega N}}}
\newcommand{\tronon}{\ensuremath{t^R_{\omega N,\omega N}}}
\newcommand{\Tpnpn}{\ensuremath{T_{\pi N,\pi N}}}
\newcommand{\Tonpn}{\ensuremath{T_{\omega N,\pi N}}}
\newcommand{\tonpn}{\ensuremath{t_{\omega N,\pi N}}}
\newcommand{\tronpn}{\ensuremath{t^R_{\omega N,\pi N}}}
\newcommand{\Tongn}{\ensuremath{T_{\omega N,\gamma N}}}
\newcommand{\tongn}{\ensuremath{t_{\omega N,\gamma N}}}
\newcommand{\trongn}{\ensuremath{t^R_{\omega N,\gamma N}}}
\newcommand{\vmbgn}{\ensuremath{v_{MB,\gamma N}}}
\newcommand{\vpngn}{\ensuremath{v_{\pi N,\gamma N}}}
\newcommand{\vongn}{\ensuremath{v_{\omega N,\gamma N}}}
\newcommand{\vonpn}{\ensuremath{v_{\omega N,\pi N}}}
\newcommand{\vpnpn}{\ensuremath{v_{\pi N,\pi N}}}
\newcommand{\vonon}{\ensuremath{v_{\omega N,\omega N}}}
\newcommand{\vrngn}{\ensuremath{v_{\rho N,\gamma N}}}
\newcommand{\tjtmbmb}{\ensuremath{t^{JT}_{M'B',MB}}}
\newcommand{\tjlsmngn}{\ensuremath{t^{JT}_{L'S'M'N',\lag\lN T_{N,z}}}}
\newcommand{\tjlsmbgn}{\ensuremath{t^{JT}_{LSMB,\lag \lN T_{N,z}}}}
\newcommand{\vjlsmngn}{\ensuremath{v^{JT}_{L'S'M'N',\lag \lN T_{N,z}}}}
\newcommand{\vjlsmbgn}{\ensuremath{v^{JT}_{LSMB,\lag \lN T_{N,z}}}}
\newcommand{\tjlsmnmb}{\ensuremath{t^{JT}_{L'S'M'N',LSMB}}}
\newcommand{\Tjlsmbmb}{\ensuremath{T^{JT}_{LSMB,L'S'M'B'}}}
\newcommand{\tjlsmbmb}{\ensuremath{t^{JT}_{LSMB,L'S'M'B'}}}
\newcommand{\tjlsmnpn}{\ensuremath{t^{JT}_{L'S'M'N',\ell \pi N}}}
\newcommand{\tjlsmbpn}{\ensuremath{t^{JT}_{LSMB,\ell \pi N}}}
\newcommand{\vjlsmnpn}{\ensuremath{v^{JT}_{L'S'M'N',\ell \pi N}}}
\newcommand{\vjlsmnmb}{\ensuremath{v^{JT}_{L'S'M'N',LSMB}}}
\newcommand{\vjlsmbpn}{\ensuremath{v^{JT}_{LSMB,\ell \pi N}}}
\newcommand{\Tjlsmngn}{\ensuremath{t^{R,JT}_{L'S'M'N',\lag\lN T_{N,z}}}}
\newcommand{\Tjlsmbgn}{\ensuremath{t^{R,JT}_{LSMB,\lag \lN T_{N,z}}}}
\newcommand{\Tfjlsmbgn}{\ensuremath{T^{JT}_{LSMB,\lag \lN T_{N,z}}}}
\newcommand{\Tjlsmnmb}{\ensuremath{t^{R,JT}_{L'S'M'N',LSMB}}}
\newcommand{\Tjlsmnpn}{\ensuremath{t^{R,JT}_{L'S'M'N',\ell \pi N}}}
\newcommand{\Tjlsmbpn}{\ensuremath{t^{R,JT}_{LSMB,\ell \pi N}}}
\newcommand{\Gbjlsi}{\ensuremath{{\Gamma}^{JT}_{LSMB,N^*_i}}}
\newcommand{\Gbjlspi}{\ensuremath{{\Gamma}^{JT}_{L'S'M'B',N^*_i}}}
\newcommand{\Gjlsi}{\ensuremath{\overline{\Gamma}^{JT}_{LSMB,N^*_i}}}
\newcommand{\Gijls}{\ensuremath{\overline{\Gamma}^{JT}_{N^*_i,LSMB}}}
\newcommand{\Gbijls}{\ensuremath{{\Gamma}^{JT}_{N^*_i,LSMB}}}
\newcommand{\Gjpn}{\ensuremath{\overline{\Gamma}^{JT}_{N^*_j,\ell\pn}}}
\newcommand{\Gign}{\ensuremath{\overline{\Gamma}^{JT}_{N^*_i,\lag\lN T_{N,z}}}}
\newcommand{\Gbign}{\ensuremath{{\Gamma}^{JT}_{N^*_i,\lag\lN T_{N,z}}}}
\newcommand{\Gjlsj}{\ensuremath{\overline{\Gamma}^{JT}_{LSMB,N^*_j}}}
\newcommand{\Gjem}{\ensuremath{\overline{\Gamma}^{JT}_{N^*_j,\lag\lN T_{N,z}}}}
\newcommand{\Ljtlsmbn}{\ensuremath{\Lambda^{JT}_{N^*LSMB}}}
\newcommand{\Drij}{\ensuremath{\mathcal{D}^{-1}_{ij}}}
\newcommand{\Mbres}{\ensuremath{M^{(0)}_{N^*}}}
\newcommand{\Cjtnlsmb}{\ensuremath{C^{JT}_{N^*LSMB}}}
\newcommand{\Ljtnlsmb}{\ensuremath{\Lambda^{JT}_{N^*LSMB}}}
\newcommand{\knstar}{\ensuremath{k_{N^*}}}
\newcommand{\vonen}{\ensuremath{v_{\omega N,\eta N}}}
\newcommand{\vonpd}{\ensuremath{v_{\omega N,\pi\Delta}}}
\newcommand{\vonsn}{\ensuremath{v_{\omega N,\sigma N}}}
\newcommand{\vonrn}{\ensuremath{v_{\omega N,\rho N}}}
\newcommand{\gnon}{\ensuremath{\gamma N\to \omega N}}
\newcommand{\gnpn}{\ensuremath{\gamma N\to \pi N}}
\newcommand{\gnky}{\ensuremath{\gamma N\to KY}}
\newcommand{\enepn}{\ensuremath{e N\to e'\pi N}}
\newcommand{\gnen}{\ensuremath{\gamma N\to \eta N}}
\newcommand{\gpop}{\ensuremath{\gamma p\to \omega p}}
\newcommand{\gpep}{\ensuremath{\gamma p\to \eta p}}
\newcommand{\gpepp}{\ensuremath{\gamma p\to \eta' p}}
\newcommand{\gnten}{\ensuremath{\gamma n\to \eta n}}
\newcommand{\pzp}{\ensuremath{\pi^0 p}}
\newcommand{\ppln}{\ensuremath{\pi^+ n}}
\newcommand{\pmp}{\ensuremath{\pi^- p}}
\newcommand{\pzn}{\ensuremath{\pi^0 n}}
\newcommand{\gppzp}{\ensuremath{\gamma p\to \pi^0 p}}
\newcommand{\gpppn}{\ensuremath{\gamma p\to \pi^+ n}}
\newcommand{\gnpmp}{\ensuremath{\gamma n\to \pi^- p}}
\newcommand{\gnpzn}{\ensuremath{\gamma n\to \pi^0 n}}
\newcommand{\gppzep}{\ensuremath{\gamma p\to \pi^0 \eta p}}
\newcommand{\pnen}{\ensuremath{\pi N\to \eta N}}
\newcommand{\pnon}{\ensuremath{\pi N\to \omega N}}
\newcommand{\pnmb}{\ensuremath{\pi N\to MB}}
\newcommand{\gnmb}{\ensuremath{\gamma N\to M\!B}}
\newcommand{\onon}{\ensuremath{\omega N\to \omega N}}
\newcommand{\pmpon}{\ensuremath{\pi^- p\to \omega n}}
\newcommand{\pnpn}{\ensuremath{\pi N\to \pi N}}
\newcommand{\pnppn}{\ensuremath{\pi N\to \pi \pi N}}
\newcommand{\knkn}{\ensuremath{K N\to K N}}
\newcommand{\nnnn}{\ensuremath{N N\to N N}}
\newcommand{\pDpD}{\ensuremath{\pi D\to \pi D}}
\newcommand{\pDpp}{\ensuremath{\pi^+ D\to pp}}
\newcommand{\Gon}{\ensuremath{G_{0,\omega N}}}
\newcommand{\Gpn}{\ensuremath{G_{0,\pi N}}}
\newcommand{\rhomb}{\ensuremath{\rho_{MB}}}
\newcommand{\rhoon}{\ensuremath{\rho_{\omega N}}}
\newcommand{\rhopn}{\ensuremath{\rho_{\pi N}}}
\newcommand{\kon}{\ensuremath{k_{\omega N}}}
\newcommand{\kpn}{\ensuremath{k_{\pi N}}}
\newcommand{\Gmb}{\ensuremath{G_{0,MB}}}
\newcommand{\Tmbgn}{\ensuremath{T_{MB,\gamma N}}}
\newcommand{\vmbpgn}{\ensuremath{v_{M'B',\gamma N}}}
\newcommand{\pntpn}{\ensuremath{\pi N\!\to\!\pi N}}
\newcommand{\pnten}{\ensuremath{\pi N\!\to\!\eta N}}
\newcommand{\pnton}{\ensuremath{\pi N\!\to\!\omega N}}
\newcommand{\epos}{\ensuremath{\slashed{\epsilon}_{\lambda_\omega}}}
\newcommand{\epo}{\ensuremath{{\epsilon}_{\lambda_\omega}}}
\newcommand{\elevi}{\ensuremath{{\epsilon}_{\alpha\beta\gamma\delta}}}
\newcommand{\eps}{\ensuremath{\epsilon}}
\newcommand{\krho}{\ensuremath{\kappa_\rho}}
\newcommand{\komg}{\ensuremath{\kappa_\omega}}
\newcommand{\komga}{\ensuremath{\kappa^t_\omega}}
\newcommand{\doh}{\ensuremath{d^{(\half)}_{\lambda'\lambda}}}
\newcommand{\dohm}{\ensuremath{d^{(\half)}_{-\lambda,-\lambda'}}}
\newcommand{\dohmo}{\ensuremath{d^{(\half)}_{\lambda',-\half}}}
\newcommand{\dohpo}{\ensuremath{d^{(\half)}_{\lambda',+\half}}}
\newcommand{\Lor}[2]{\ensuremath{\Lambda^{#1}_{\hphantom{#1}{#2}}}}
\newcommand{\ILor}[2]{\ensuremath{\Lambda_{#1}^{\hphantom{#1}{#2}}}}
\newcommand{\LorT}[2]{\ensuremath{[\Lambda^T]^{#1}_{\hphantom{#1}{#2}}}}
\newcommand{\dsdo}{{\frac{d\sigma}{d\Omega}}}
\newcommand{\dspdo}{\ensuremath{{\frac{d\sigma_\pi}{d\Omega}}}}
\newcommand{\dsgdo}{\ensuremath{{\frac{d\sigma_\gamma}{d\Omega}}}}
\newcommand{\chipd}{\ensuremath{\chi^2/N_d}}
\newcommand{\chipda}{\ensuremath{\chi^2(\alpha)/N_d}}
\newcommand{\bpop}{\ensuremath{\bvec{p}'_1}}
\newcommand{\bptp}{\ensuremath{\bvec{p}'_2}}
\newcommand{\bpip}{\ensuremath{\bvec{p}'_i}}
\newcommand{\bpo}{\ensuremath{\bvec{p}_1}}
\newcommand{\bpt}{\ensuremath{\bvec{p}_2}}
\newcommand{\bpi}{\ensuremath{\bvec{p}_i}}
\newcommand{\bqo}{\ensuremath{\bvec{q}_1}}
\newcommand{\bqt}{\ensuremath{\bvec{q}_2}}
\newcommand{\bqi}{\ensuremath{\bvec{q}_i}}
\newcommand{\bQ}{\ensuremath{\bvec{Q}}}
\newcommand{\bq}{\ensuremath{\bvec{q}}}
\newcommand{\ketq}{\ensuremath{\ket{\bqo,\bqt}}}
\newcommand{\ketqc}{\ensuremath{\ket{\bQ,\bq}}}
\newcommand{\bP}{\ensuremath{\bvec{P}}}
\newcommand{\bPp}{\ensuremath{\bvec{P}'}}
\newcommand{\bpr}{\ensuremath{\bvec{p}}}
\newcommand{\bprp}{\ensuremath{\bvec{p}'}}
\newcommand{\ketPsiq}{\ensuremath{\ket{\Psi_{\bq}^{(\pm)}}}}
\newcommand{\ketPsiqQ}{\ensuremath{\ket{\Psi_{\bQ,\bq}^{(\pm)}}}}
\newcommand{\Ld}{\ensuremath{\mathcal{L}}}
\newcommand{\ps}{\mbox{ps}}
\newcommand{\fndp}{f_{N\Delta\pi}}
\newcommand{\fndr}{f_{N\Delta\rho}}
\newcommand{\said}{{\sc said}}
\newcommand{\ret}{\ensuremath{\langle{\tt ret}\rangle}}
\newcommand{\ddf}[1]{\ensuremath{\delta^{(#1)}}}
\newcommand{\Tpp}{\ensuremath{T_{\pi\pi}}}
\newcommand{\Kpp}{\ensuremath{K_{\pi\pi}}}
\newcommand{\Tpe}{\ensuremath{T_{\pi\eta}}}
\newcommand{\Kpe}{\ensuremath{K_{\pi\eta}}}
\newcommand{\Tep}{\ensuremath{T_{\eta\pi}}}
\newcommand{\Kep}{\ensuremath{K_{\eta\pi}}}
\newcommand{\Tee}{\ensuremath{T_{\eta\eta}}}
\newcommand{\Kee}{\ensuremath{K_{\eta\eta}}}
\newcommand{\Tpig}{\ensuremath{T_{\pi\gamma}}}
\newcommand{\Kpig}{\ensuremath{K_{\pi\gamma}}}
\newcommand{\oKpig}{\ensuremath{\overline{K}_{\pi\gamma}}}
\newcommand{\tKpig}{\ensuremath{\tilde{K}_{\pi\gamma}}}
\newcommand{\Teg}{\ensuremath{T_{\eta\gamma}}}
\newcommand{\Keg}{\ensuremath{K_{\eta\gamma}}}
\newcommand{\Kab}{\ensuremath{K_{\alpha\beta}}}
\newcommand{\R}{\ensuremath{\mathbb{R}}}
\newcommand{\C}{\ensuremath{\mathbb{C}}}
\newcommand{\Ezp}{\ensuremath{E^{\pi}_{0+}}}
\newcommand{\Eze}{\ensuremath{E^{\eta}_{0+}}}
\newcommand{\Ga}{\ensuremath{\Gamma_\alpha}}
\newcommand{\Gb}{\ensuremath{\Gamma_\beta}}
\newcommand{\RH}{\ensuremath{\mathcal{R}\!\!-\!\!\mathcal{H}}}
\newcommand{\calT}{\mathcal{T}}
\newcommand{\maid}{{\sc maid}}
\newcommand{\Kbar}{\ensuremath{\overline{K}}}
\newcommand{\zbar}{\ensuremath{\overline{z}}}
\newcommand{\kbar}{\ensuremath{\overline{k}}}
\newcommand{\dom}{\ensuremath{\mathcal{D}}}
\newcommand{\domi}[1]{\ensuremath{\mathcal{D}_{#1}}}
\newcommand{\pbar}{\ensuremath{\overline{p}}}
\newcommand{\Nab}{\ensuremath{N_{\alpha\beta}}}
\newcommand{\Nee}{\ensuremath{N_{\eta\eta}}}
\newcommand{\dth}[1]{\delta^{(3)}(#1)}
\newcommand{\dfo}[1]{\delta^{(4)}(#1)}
\newcommand{\intk}{\int\!\!\frac{d^3\! k}{(2\pi)^3}}
\newcommand{\intkg}{\int\!\!{d^3\! k_\gamma}}
\newcommand{\intks}{\int\!\!{d^3\! k_\sigma}}
\newcommand{\nch}{\ensuremath{N_{\mbox{ch}}}}
\newcommand{\nc}{\ensuremath{N_{ch}}}
\newcommand{\re}{\ensuremath{\mbox{Re }\!}}
\newcommand{\im}{\ensuremath{\mbox{Im }\!}}
\newcommand{\EetaS}{\ensuremath{E^\eta_{0+}}}
\newcommand{\EpiS}{\ensuremath{E^\pi_{0+}}}
\newcommand{\tobull}{\ensuremath{\to}}
\newcommand{\Kcm}{\ensuremath{K_{CM}}}
\newcommand{\lra}{\ensuremath{\leftrightarrow}}

\newcommand{\gn}{\ensuremath{\gamma N}}
\newcommand{\gp}{\ensuremath{\gamma p}}
\newcommand{\geta}{\ensuremath{\gamma \eta}}
\newcommand{\pp}{\ensuremath{pp}}
\newcommand{\pn}{\ensuremath{\pi N}}
\newcommand{\phn}{\ensuremath{\pi d}}
\newcommand{\en}{\ensuremath{\eta N}}
\newcommand{\epn}{\ensuremath{\eta' N}}
\newcommand{\pD}{\ensuremath{\pi \Delta}}
\newcommand{\sn}{\ensuremath{\sigma N}}
\newcommand{\rn}{\ensuremath{\rho N}}
\newcommand{\on}{\ensuremath{\omega N}}
\newcommand{\ppn}{\ensuremath{\pi\pi N}}
\newcommand{\pipi}{\ensuremath{\pi\pi}}
\newcommand{\kn}{\ensuremath{KN}}
\newcommand{\ky}{\ensuremath{KY}}
\newcommand{\kl}{\ensuremath{K\Lambda}}
\newcommand{\ks}{\ensuremath{K\Sigma}}
\newcommand{\bn}{\ensuremath{eN}}
\newcommand{\pR}{\ensuremath{\pi N^*}}
\newcommand{\bpn}{\ensuremath{e\pi N}}
\newcommand{\fpo}{\ensuremath{5\oplus 1}}
\newcommand{\faoe}{{\sc FA08}}
\newcommand{\fpoe}{{\sc FP08}}
\newcommand{\fsoe}{{\sc FS08}}
\newcommand{\psic}{\ensuremath{\psi_{n\kappa jm}}}
\newcommand{\hi}[1]{\ensuremath{H_I(t_{#1})}}
\newcommand{\dcp}{\ensuremath{\mathcal{D}}}
\newcommand{\pptopp}{\ensuremath{\pipi\to\pipi}}
\newcommand{\pntoppn}{\ensuremath{\pn\to\ppn}}
\newcommand{\ovlt}{\ensuremath{\overline{t}}}
\newcommand{\smat}{\ensuremath{e^{-i\int_{-\infty}^\infty\!dt\,H_I(t)}}}

\newcommand{\itPFP}{\textit{Physics for Future Presidents}}
\newcommand{\itaPFP}{\textit{PFP}}
\newcommand{\prle}{\textit{PR}\textbf{97}}

\title{Functional reduction of the $S$ matrix in the canonical formalism}
\author{Mark W.\ Paris}
\email{mparis@gwu.edu}
\affiliation{Data Analysis Center, Institute for Nuclear Studies, The
George Washington University\\
725 21$^{\mbox{st}}$ Street NW, Washington, DC 20052}

\date{\today}

\begin{abstract}
The Low equation is derived in a functional approach to the reduction
of the $S$ matrix in the canonical formalism. This establishes the
vacuum expectation value of the scattering matrix as the generating
functional of non-forward Green functions, without reference to
external currents. The method provides an alternate derivation of
non-perturbative results of field theory, such as the Low equation,
and considerably simplifies their derivation as well as that of the
rules of perturbation theory, the LSZ reduction formula, the
Dyson-Schwinger equations and crossing symmetry. The approach is
employed to further develop the Low equation via reduction of the
fermionic sector to obtain a reduced Dyson-Schwinger equation for
boson-fermion scattering.
\end{abstract}

\maketitle

Functional methods in field theory have proven to be useful in the
development of an array of non-perturbative and perturbative results.
The path integral approach, formulated through an invariant Lagrangian
functional of $c$-number fields and external currents, constitutes the
canonical example of this success. The complementary, and at least
empirically equivalent, approach of the canonical quantization of a
classical field theory is based on the algebra of annihilation
operators and their Hermitian adjoints. The purpose of this 
\textit{Letter} is to develop a
functional approach to the operator algebra within the canonical
formalism, an immediate consequence of which is the identification of
the vacuum expectation value of the $S$ matrix as the generating
functional of the scattering and reaction amplitudes of the theory. In
addition, the interpretation of the field operators as non-commuting
operators may be abandoned. Only time ordering and the Grassmanian
algebraic properties of fermionic fields need be taken into account.

Functional methods have been employed in the canonical formulation in
previous works. Schweber\cite{Schweber:1955yf} considered a
functional-derivative representation of the field commutator in a
derivation of the scattering matrix elements in terms of Heisenberg
field operators in the K\"{a}ll\'{e}n-Yang-Feldman formalism.  The
present study may be viewed as an extension of the
functional-derivative representation of the commutator of
Ref.\cite{Schweber:1955yf}. Functional representations of the $S$
matrix have also been considered extensively by Rzewuski and
Garbaczewski\cite{Rzewuski:1970ef,*Garbaczewski:1974gf,*RGnote} in the
study of convergence and asymptotic behavior of field theory, and by
Efimov\cite{Efimov:1975sm}, in the proof of $S$-matrix unitarity.  The
functional approach of determining the relationship between
time-ordered and normal ordered products given by Akhiezer and
Berestetski\u\i\cite{Akhiezer:1965qe} is closely related to that given
here with rather a different emphasis.  Their method has focused, in
particular, on the determination of the rules of perturbation theory.

A classic example of a non-perturbative result in quantum field theory
is the Low equation [Eq.\eqref{eqn:df2Sdada}] for boson-fermion
scattering\cite{Low:1955PhRv}.  The matrix element for meson-nucleon
scattering is
\begin{align}
\label{eqn:Sbb}
S_{fi} &= \bra{p'} a_j(q') S a_i^\dag(q) \ket{p},
\end{align}
where $p (p')$ is the nucleon initial (final) three-momentum, $q(q')$
is the meson initial (final) three-momentum, $i(j)$ is the
three-component of the meson isospin, $i,j = \pm,0$, and $a_i(q)$ is
the annihilation operator. The subscript $i(f)$ is short-hand for the
initial (final) states $pqi(p'q'j)$, suppressing the nucleon spin.  The
scattering operator $S$ in the Dirac (interaction) representation is
\begin{align}
\label{eqn:UtoSexp}
S &= P e^{-i\int_{-\infty}^\infty\! dt\, H_I(t)},
\end{align}
where $H_I$ is the interaction Hamiltonian and $P$ is the Dyson
chronological product (DCP). 

The standard reduction of the bosonic sector of the $S$ matrix
evaluates the commutators
\begin{align}
\label{eqn:Sfi}
S_{fi} &= \delta_{ij} \delta^{(3)}(q-q') \bra{p'}S\ket{p} \nonumber \\
&+\bra{p'}\comm{a_j(q')}{\comm{S}{a_i^\dag(q)}}\ket{p},
\end{align}
and employs the Dyson series for the $S$ matrix\cite{Dyson:1949bp} by
repeated application of the commutator identity $\comm{ab}{c} =
a\comm{b}{c} + \comm{a}{c}b$. In the present work, we replace the
manipulations of the annihilation operator algebra with those of
functional differentiation. In so doing, we exploit an isomorphism of
the annihilation operator algebra with that of a \textit{derivation}
over functionals of annihilation operators and their adjoints. This is
accomplished through the following relation:
\begin{align} 
\label{eqn:dF} 
\comm{F[a,a^\dag]}{a^\dag_i(q)} &=
\frac{\delta F[a,a^\dag]}{\delta a_i(q)}, 
\end{align} 
and it's Hermitian conjugate. The application of Eq.\eqref{eqn:dF} to
the commutators in Eq.\eqref{eqn:Sfi} constitutes the functional
reduction of the $S$ matrix in the canonical formalism.  We emphasize
that, in contrast to the path integral formulation of field theory,
the functional derivatives in the canonical approach considered here
are of the $S$ matrix (or its vacuum expectation value) and with
respect to the fields. No classical, external currents or fields have
been introduced.

A particular example of the use of this relation is afforded by
letting $F$ be the isospin $i$ meson configuration space field
operator, $\phi_i(x)$:
\begin{align}
\label{eqn:funca}
\fnfd{\phi_i(x)}{a_j(q)}
&= \delta_{ji} \frac{e^{-iq\cdot x}}{(2\pi)^{3/2}(2\omega_q)^{1/2}}
= \comm{\phi_i(x)}{a_j^\dag(q)}.
\end{align}
This relation establishes the isomorphism of the commutator to the
functional derivative, reproducing Eqs.(1.3) and (1.4) from
Ref.\cite{Low:1955PhRv}, although with a different normalization for
the pion wave functions.  

Replacing the operator algebra by the functional calculus, for
functionals of both fermionic (see below) and bosonic fields, such as
the $S$ matrix, is useful for the simplification of the derivation of
non-perturbative results of canonical field theory -- the Low
equation\cite{Low:1955PhRv}, for example.  This technique offers an
alternative to either the annihilation operator calculus or the
path-integral approach. It permits the derivation of the Low equation,
the Feynman rules, the Dyson-Schwinger equations, and other
established results with a significant reduction of work and
complexity, compared to the operator algebra approach.  Additionally,
we see the $S$ matrix in a new light.  Generally, the
in-vacuum-to-out-vacuum amplitude is the generating functional for all
scattering processes involving on- or off-shell particles, in any
combination (see Eqs.\eqref{eqn:f2S-gen} and \eqref{eqn:0S0} below).

As a specific application of the functional approach in the canonical
formulation of field theory we consider the determination of the
non-forward scattering matrix element for meson-nucleon scattering. We
first consider just the reduction of the $S$ matrix in the bosonic
sector. Employing Eq.\eqref{eqn:dF} to write Eq.\eqref{eqn:Sbb} in
terms of the functional derivatives with respect to the annihilation
operators and their adjoints, we obtain
\begin{align}
\label{eqn:sbb1}
a_j(q') S a^\dag_i(q) &= a^\dag_i(q) S a_j(q')
 +\delta_{ij}\delta^{(3)}(q'-q)S \nonumber \\
&+\fnfd{S}{a_j(q')} a_i 
 +\fdfd{a^\dag_j(q')}\fnfd{S}{a_i(q)}.
\end{align}
Neglecting terms that vanish upon taking the expectation value of this
expression with respect to the meson vacuum gives
\begin{align}
\label{eqn:f2S}
\fdfd{a^\dag_j(q')}\fnfd{S}{a_i(q)}
&= a_j(q') S a^\dag_i(q)-\delta_{ij}\delta^{(3)}(q'-q)S.
\end{align}
This result will be useful in the derivation of the Low equation.
Comparison of Eq.\eqref{eqn:Sfi} with Eq.\eqref{eqn:f2S} demonstrates
the relationship between the nested commutator and the second
functional derivative with respect to the annihilation operator and
its adjoint. We note that for the generic functional $F=F[a,a^\dag]$,
the symmetry of the order of differentiation
\begin{align}
\label{eqn:fdsymm}
\fdfd{a^\dag_j(q')}\fnfd{F}{a_i(q)} &=
\fdfd{a_i(q)}\fnfd{F}{a^\dag_j(q')},
\end{align}
is ensured by the Jacobi identity and
$\comm{a_i(q)}{a^\dag_j(q')} = \delta_{ij}\delta^{(3)}(q-q')$.

Equation \eqref{eqn:f2S} is the result that functional derivatives of
$S$ with respect to the annihilation fields and their adjoints yields
the non-forward $S$-matrix element (upon calculating the above
operators' expectation value in the meson vacuum). This is reminiscent of
the familiar result in the Lagrangian, path integral formulation
wherein functional derivatives of a generating functional with respect
to classical, external currents yield the scattering and reaction
amplitudes.  Here we have derived a functional result within the
canonical formalism of field theory.

Using Eq.\eqref{eqn:f2S} we turn to the derivation of the
Low equation, Eq.(1.9) in Ref.\cite{Low:1955PhRv}. We require
the calculation of the functional derivatives of $S$
[Eq.\eqref{eqn:UtoSexp}]:
\begin{align}
\label{eqn:dSda}
\frac{\delta S}{\delta a_i(q)}
&= -iP \left\{
e^{-i\int_{-\infty}^\infty dt H_I(t)}
\int_{-\infty}^\infty dt 
\fnfd{H_I(t)}{a_i(q)} \right\}.
\end{align}
We have exploited the property of the DCP that allows us to disregard
the noncommutativity of its arguments in the above relation. We might
note in passing that the DCP gives a representation of the operator
ordering calculus of Feynman\cite{Feynman:1951gn}.

The following results can be derived independently of the specific
form of the interaction Hamiltonian functional, $H_I(t)$, as discussed
below. However, to be concrete, we consider the form taken by
Low\cite{Low:1955PhRv}:
\begin{align}
\label{eqn:HI}
\mathcal{H}_I(x) &= ig_0 \dcpsi(x) \gamma_5 \tau_i \psi(x) \phi_i(x)
+ \sqtr\lambda [\phi_i(x) \phi_i(x)]^2 \nonumber \\
& - \delta m \dcpsi(x)\psi(x)
 - \shalf \delta\mu^2 \phi_i(x) \phi_i(x),
\end{align}
with $x \equiv (t,\xvec)$ and $H_I(t)=\int\,d^3\!x\,\mathcal{H}_I(x)$.
The functional chain rule is employed as:
\begin{align}
\label{eqn:funcH}
\fnfd{\hi{\null}}{a_i(q)}
&= \int\! d^4\!x' \,
\fnfd{\phi_k(x')}{a_i(q)}\fnfd{\hi{\null}}{\phi_k(x')},
\end{align}
where a $\sum_k$ is implied. Using the above relations, we obtain
\begin{align}
\label{eqn:fdHda}
\fnfd{\hi{\null}}{a_i(q)}
&= \int\! d^3\!x \, 
\frac{e^{-iq\cdot x}}{(2\pi)^{3/2}(2\omega_q)^{1/2}}J_i(x), \\
J_i(x) &= ig_0\dcpsi(x)\gamma_5\tau_i\psi(x)
- \delta\mu^2 \phi_i(x) \nonumber \\
&+ \lambda \phi_j(x) \phi_j(x) \phi_i(x).
\end{align}


It is straightforward from this point to evaluate the functional
derivatives of $S$ in Eq.\eqref{eqn:f2S} required for the
meson-nucleon scattering matrix element. We have
\begin{align}
\label{eqn:df2Sdada}
&\fdfd{a_j^\dag(q')}\fnfd{S}{a_i(q)} \nonumber \\
&= (-i)^2\int\! {d^4\!x' d^4\!x}
\frac{e^{iq'\cdot x'}e^{-iq\cdot x}}
{(2\pi)^3\sqrt{4\omega_{q'}\omega_q}}
(\partial_{x'}^2 + \mu^2)(\partial_{x}^2 + \mu^2) \nonumber \\
&\times P[\Omega^\dag(\infty)\bfphi_j(x')\bfphi_i(x)\Omega(-\infty)].
\end{align}
where $\Omega(t)=e^{i(H_0+H_I)t}e^{-iH_0t}$.  In obtaining
Eq.\eqref{eqn:df2Sdada}, an operator relation, we have ignored
disconnected terms and those that have vanishing vacuum expectation
value, used
\begin{align}
\label{eqn:dfOdadag}
\fnfd{J_i(x)}{\phi_j(x')} = \delta^{(4)}(x-x') \big[
&-\delta\mu^2+\lambda(2\phi_k(x')\phi_i(x')\nonumber \\
&+\delta_{ki}\phi_j(x')\phi_j(x'))
\big],
\end{align}
the boson field equation of motion,
$(\partial^2+\mu^2)\bfphi_i(x)=-\bfJ_i(x)$, and the relation between
operators, $\bm{O}(x) = \Omega(x^0) O(x) \Omega^\dag(x^0)$ in the
Heisenberg and Dirac representations; Heisenberg operators are denoted
in bold-type. Taking the vacuum expectation value of
Eq.\eqref{eqn:df2Sdada} and using the relationship between the in- and
out-vacuum states, $\Psi^\pm$ and the Dirac representation vacuum,
$\Phi$, $\Psi^{\pm}=\Omega(\mp\infty)\Phi$ gives the main result of
Ref.\cite{Low:1955PhRv}, Eq.(1.9), the Low equation.  It also provides
an independent derivation of the LSZ reduction
formula\cite{Lehmann:1957zz}. Note that despite the choice of a
specific form for the interaction Hamiltonian [Eq.\eqref{eqn:HI}] the
result of Eq.\eqref{eqn:df2Sdada} is independent of the particular
local, field-theoretic model assumed\cite{Haag:1957rs}.

The fermionic sector is amenable to a generalization of the above
procedure.  Defining left- and right-derivatives with respect to
fermion (antifermion) annihilation operators and adjoints
$b_s(p),b^\dag_s(p)$ $(d_s(p),d^\dag_s(p))$ for spin-$s$, and
momentum $p$, according to the Grassmann algebra, we obtain:
\begin{align}
\label{eqn:f2S-fermi}
\fdfd{b^\dag_{s'}(p')}S\rfdfd{b_s(p)}
&= b_s(p') S b^\dag_s(p)-\delta_{s's}\delta^{(3)}(p'-p)S.
\end{align}
The fermionic derivatives commute with the bosonic derivatives in
Eq.\eqref{eqn:df2Sdada} and we obtain for the meson-nucleon scattering
amplitude
\begin{align}
\label{eqn:bpapSba}
&\fdfd{b^\dag_{s'}(p')}\fdfd{a_j^\dag(q')}S\rfdfd{a_i(q)}\rfdfd{b_s(p)}
\nonumber \\
&= (-i)^2\int\! {d^4\!x' d^4\!x} 
\frac{e^{iq'\cdot x'}e^{-iq\cdot x}}
{(2\pi)^6\sqrt{4\omega_{q'}\omega_q}}\frac{m}{\sqrt{E_{p'}E_p}}
\Omega^\dag(\infty)\nonumber \\
&\times\Bigg\{  \overline{u}_{p's'} \Big\{
\big[(-ig_0)^2
\gamma_5\tau_j T(\bfpsi(x')\dcbfpsi(x))\gamma_5 \tau_i
e^{ip'\cdot x'}e^{ip\cdot x}
\nonumber \\
&-ig_0\gamma_5\tau_j\!\int\! d^4\!x_1
T(\bfpsi(x')\bfJ_i(x)\overline{\bfj}(x_1))
e^{ip'\cdot x}e^{-ip\cdot x_1}
\nonumber \\
\displaybreak
&-ig_0\!\int\! d^4\!x_1
T(\bfj(x_1)\bfJ_j(x')\dcbfpsi(x))
\gamma_5\tau_i
e^{ip'\cdot x_1}e^{-ip\cdot x} \big] \nonumber \\
&+\big[x'\!j\leftrightarrow xi\big] \Big\}u_{ps}
-iT\Big\{\bfJ_j(x')\bfJ_i(x)
\int\! {d^4\!x_1 d^4\!x_2}  \nonumber \\ &\times
\overline{u}_{p's'}
\big[-i\bfj(x_1)\overline{\bfj}(x_2)
+\delta^{(4)}(x_1-x_2) \nonumber \\ &\times\gamma_5\tau_k\bfphi_k(x_1)
e^{i(p'-p)\cdot x_1}\big]
u_{ps}\Big\}\Bigg\}\Omega(-\infty),
\end{align}
where we have neglected terms that have vanishing vacuum expectation
value as well as disconnected terms and introduced, as usual, the
time-ordering operator $T$ for Dirac fields, which coincides with the
DCP for bosonic operators. 
In
Eq.\eqref{eqn:bpapSba}, we have also defined the current, $\bfj(x)$ as
the inhomogeneous term in the nucleon field operator equation of
motion, $(i\slashed{\partial}-m)\bfpsi(x)=\bfj(x)$. Upon taking the
vacuum expectation value of the above expression we recover the
reduced Dyson-Schwinger equation for \pntpn\
scattering\cite{Dyson:1949ha,*Schwinger:1951ex,*Schwinger:1951hq}.
The expression above can be simplified further by employing the
Heisenberg equations of motion of the field operators, which would
yield the bosonic and fermionic LSZ reduction of the \pntpn\ $S$
matrix. Crossing symmetry follows immediately in this approach since,
for example, both $\delta/\delta b_s(p)$ and $\delta/\delta
d^\dag_s(p)$ are proportional to $\delta/\delta\psi(x)$.

It is useful to consider each term in the above expression in terms of
its perturbation expansion. The first term is the exact fermion
propagator taken between bare vertices. It may be expanded in a series
of one-particle irreducible graphs in the usual way and corresponds to
$s$-channel structure of the amplitude, partly determining resonance
poles and threshold branch points; the corresponding meson-crossed
term yields $u$-channel structure.  The following two terms correspond
to diagrams which have bare couplings at one vertex and three-point
functions at the other vertex.  These three-point functions, formed of
products like $\bfj\bfJ\dcbfpsi{}$, correspond to vertex and external
particle renormalization in the perturbation theory. The next three
exchange terms, $[x'j\leftrightarrow xi]$ are the crossed-meson
contributions, as mentioned.  The penultimate term, proportional to
$\bfJ\bfJ\bfj\bfj$, corresponds to graphs with at least two meson
vertices on external fermion lines, corresponding to internal
meson propagators. The last term gives, at lowest order, the one-loop,
fermion triangle graph.

In addition to simplifying the derivation of non-perturbative results
within canonical field theory the functional reduction of the $S$
matrix may be applied, with significant pedagogical value, to the
problem of deriving the rules of perturbation theory. As an
illustration, we continue to work with the \pn\ system with
interaction described by Eq.\eqref{eqn:HI}. We consider the matrix
element for \pn\ scattering
\begin{align}
\label{eqn:pt-pnpn}
\fdfd{b^\dag_{s'}(p')}\fdfd{a_j^\dag(q')}S^{(2)}\rfdfd{a_i(q)}\rfdfd{b_s(p)},
\end{align}
where $S^{(2)}$ is $\frac{(-i)^2}{2}\int dt_1 dt_2
P[H_I(t_1)H_I(t_2)]$. (The zeroth and first-order $S$ make no
contribution to the \pntpn\ process since there are four functional
derivatives.) Using the identity
\begin{align}
\label{eqn:Pord-T}
&\fdfd{b_{s'}(p')}P[J_j(x')J_i(x)]\rfdfd{b_{s}(p)}
\nonumber \\
&= T\left(\fnfd{J_j(x')}{b^\dag_{s'}(p')}\rfnfd{J_i(x)}{b_{s'}(p')}\right)
+ T\left(\fnfd{J_i(x)}{b^\dag_{s'}(p')}\rfnfd{J_j(x')}{b_{s'}(p')}\right),
\end{align}
where $\rfnfd{}{b_{s'}(p')}$ is the right-derivative, one obtains the
standard $s$- and $u$-channel nucleon exchange tree-level amplitudes,
which correspond to the first and second terms of
Eq.\eqref{eqn:bpapSba}, at leading order in the bare coupling
constant, $g_0$. Equation \eqref{eqn:Pord-T} is readily generalized to
the higher $n$-point functions where $n>2$\cite{Paris:2011ip}.

Generalizing to processes involving $n$ incident and $m$ final
asymptotic state particles, one arrives at the relationship between
the Heisenberg vacuum-to-vacuum transition element,
$(\Psi^{(-)}_0,\Psi^{(+)}_0)=(\Phi_0,S\Phi_0)$ and the non-diagonal
(ND) $S$-matrix elements:
\begin{align}
\label{eqn:f2S-gen}
&\ovl{p_1'\cdots p_m';\mbox{out}}{p_1\cdots p_n;\mbox{in}}
\underset{\mbox{\tiny ND}}{=} \nonumber \\
&\fdfd{c_1^\dag(p'_1)}\cdots \fdfd{c_m^\dag(p'_m)}
(\Phi_0,S\Phi_0)
\rfdfd{c_{m+1}(p_1)}\cdots \rfdfd{c_{m+n}(p_n)},
\end{align}
where $c_i$, $i=1,\ldots,m+n$ corresponds to the annihilation
operator appropriate to the relevant particle or antiparticle
single-particle states. A consequence of the above relation is that
the vacuum expectation value $S$-matrix functional $(\Phi_0,S\Phi)$
may be represented as
\begin{align}
\label{eqn:0S0}
(\Phi_0,S\Phi_0) &= \sum_{m,n}
\int d^3p'_1\cdots d^3p'_m\int d^3p_1\cdots d^3p_n\nonumber\\
&c_1 \cdots c_m \mathcal{M}_{mn} c^\dag_{m+1} \cdots c^\dag_{m+n},
\end{align}
where $\mathcal{M}_{mn}$ is the non-forward scattering (or reaction)
matrix element for $n$ incident and $m$ final state particles. This
representation of the vacuum expectation value of the $S$ matrix
raises the interesting possibility of evaluating the scattering and
reaction amplitudes numerically. Importance sampling of the phase of
$(\Phi_0, S\Phi_0)$ and numerical evaluation of the functional
derivatives may offer an approach complementary to lattice studies
that evaluate correlation functions numerically.

The preceding results have all been derived by functional
differentiation with respect to the momentum eigenstate annihilation
operators and their adjoints. It is perhaps worth mentioning here that
much of the analysis can be repeated by differentiating with respect
to the configuration space field. The boson field, for example, has
\begin{align}
\label{eqn:a-phi}
\fdfd{a_i(q)} &= \int d^4\!x \fnfd{\phi_k(x)}{a_i(q)}\fdfd{\phi_k(x)},
\nonumber \\
              &= \frac{1}{(2\pi)^{3/2}}
                 \frac{1}{\sqrt{2\omega_q}}
                 \int d^4\!x e^{-iq\cdot x}\fdfd{\phi_i(x)},
\end{align}
Additionally, results off the
mass-shell may be obtained by considering the effect of functional
derivatives with respect to fields of the $n$-point functions of
the theory, rather than directly on the $S$ matrix. An example of this
is the relationship
\begin{align}
\label{eqn:d2-3}
&\fdfd{a_i(q)}T(\bfpsi(x')\dcbfpsi(x))
= \frac{1}{(2\pi)^{3/2}}
   \frac{1}{\sqrt{2\omega_q}} \nonumber \\
&\int d^4\!x_1\,e^{-iq\cdot x_1} T(\bfpsi(x')\dcbfpsi(x)\bfJ(x_1))
\end{align}
between the two-point and three-point Green functions.

The above analysis can be carried out in its entirety for any set of
fields and assumed interaction Hamiltonian including gauge fields and
fields of higher spin.

The functional reduction of the $S$ matrix has been developed as a
`third way' to derive results of fundamental importance in
non-perturbative and perturbative field theory. The approach is
equivalent in all its consequences to the operator algebra approach of
canonical field theory and complementary to the path integral
formulation. We have reproduced the Low scattering equation for
fermions and bosons, the LSZ reduction, and the Dyson-Schwinger
equations through a simplified computational technology. The
identification of the vacuum expectation value of the $S$ matrix as
the generating functional over annihilation operators and their
adjoints is established. Finally, it is hoped that the method will
have pedagogical value to establishing both the rules of perturbation
theory and non-perturbative results. This hope stems from the
familiarity of the rules of differentiation (Leibniz, chain, etc.) at
the small expense of learning the functional calculus.

The author thanks H.\ Grie{\ss}hammer and W.\ Parke for insightful
comments. This work was supported by the US Department of Energy
Grant No.\ DE-FG02-99-ER41110.

\bibliographystyle{apsrevM}
\bibliography{master}
\end{document}